\documentclass[runningheads]{llncs}
\usepackage[T1]{fontenc}
\usepackage{graphicx}
\usepackage{todonotes} 
\usepackage{siunitx} 
\usepackage[sort,compress]{cite}

\usepackage{amsmath}
\usepackage{amssymb}
\usepackage{amsfonts}
\usepackage[hyphens]{url}

\usepackage{algorithm}
\usepackage{algorithm}
\usepackage{algpseudocode}
\usepackage{tikz}
\usetikzlibrary{arrows.meta,patterns}
\usepackage{caption}
\usepackage{xcolor}
\usepackage{hyperref}
\usepackage{cleveref}

\usepackage{listings}

\lstdefinestyle{inlinecpp}{
  language=C++,
  basicstyle=\ttfamily,
  keywordstyle=\bfseries,
  commentstyle=\slshape,
  stringstyle=\ttfamily,
  identifierstyle=,
}

\lstdefinestyle{inlinesh}{
  language=sh,
  basicstyle=\ttfamily,
}

\hypersetup{
    pdfborder={0 0 0},
    colorlinks=true,
    linkcolor=black,
    citecolor=black,
    filecolor=black,
    urlcolor=blue
}

\colorlet{potrf}{purple}
\colorlet{trsm}{teal}
\definecolor{syrk}{rgb}{0.8, 0.8, 0}
\colorlet{gemm}{blue!65}
\definecolor{utile}{rgb}{0.8, 0.8, 0.8}

\begin{document}
\title{GPU-Resident Gaussian Process Regression Leveraging Asynchronous Tasks with HPX}

\titlerunning{GPU-Resident GPR Leveraging Asynchronous Tasks with HPX}
\author{Henrik Möllmann \orcidID{0009-0007-1006-5122} \and
Dirk Pflüger \orcidID{0000-0002-4360-0212}
\and Alexander Strack\orcidID{0000-0002-9939-9044}}
\authorrunning{H. Möllmann \and et al.}
%
\institute{Institute of Parallel and Distributed Systems, University of Stuttgart,\\ 70569 Stuttgart, Germany\\
\email{st178292@stud.uni-stuttgart.de}\newline
\email{\{alexander.strack, dirk.pflueger\}@ipvs.uni-stuttgart.de}
}
\maketitle              

\begin{abstract}
Gaussian processes (GPs) are a widely used regression tool, but the cubic complexity of exact solvers limits their scalability.
To address this challenge, we extend the GPRat library by incorporating a fully GPU-resident GP prediction pipeline.
GPRat is an HPX-based library that combines task-based parallelism with an intuitive Python API.

We implement tiled algorithms for the GP prediction using optimized CUDA libraries, thereby exploiting massive parallelism for linear algebra operations.
We evaluate the optimal number of CUDA streams and compare the performance of our GPU implementation to the existing CPU-based implementation.
Our results show the GPU implementation provides speedups for datasets larger than $128$ training samples. 
We observe speedups of up to $\num{4.3}$ for the Cholesky decomposition itself and $\num{4.6}$ for the GP prediction.
Furthermore, combining HPX with multiple CUDA streams allows GPRat to match, and for large datasets, surpass cuSOLVER's performance by up to \SI{11}{percent}.

\keywords{Gaussian Process Regression \and Cholesky Decomposition \and Asynchronous Many-task Runtimes \and Tiled Algorithms \and HPX \and CUDA \and GPRat}
\end{abstract}
\section{Introduction}
\label{sec:intro}

The rapid advancement of artificial intelligence (AI) drives hardware development, necessitating software that can effectively utilize specialized accelerators.
Consequently, existing libraries must adapt to support GPU computation to keep up with growing computational demands.

Gaussian Processes (GPs) are powerful probabilistic models used in diverse fields, ranging from AI~\cite{damianou2013deep} and scientific computing~\cite{kennedy2001bayesian} to big data~\cite{liu2020bigdata} and control theory~\cite{kocijan2016modelling,berkenkamp2015control}. 
One key application of GPs in control theory is system identification (SI)~\cite{ljung2010si}, specifically for modeling non-linear dynamic systems.
In this work, we consider a coupled mass-spring-damper system where the position of the last mass is non-linearly dependent on the force applied to the first mass.
We employ an NFIR model, where the system state is represented by a feature vector $\mathbf x_i \in \mathbb R^D$ containing $D$ regressors.
Using a simulator, we generate training and test datasets of input-output pairs $(\mathbf x_i, y_i)$ by observing the input force and the output position at a constant rate.

The popularity of GPs is not only linked to their ability to capture non-linear relationships but also to their ability to provide uncertainty estimates.
However, making exact predictions with or without uncertainty estimation requires the inversion of a large covariance matrix. 
This inversion can be exactly computed with the Cholesky decomposition. Thus, the cubic time complexity $\mathcal O(n^3)$ dominates the runtime.
While libraries like GPyTorch~\cite{gardner2018gpytorch} and GPflow~\cite{matthews2017gpflow} address this challenge by relying on the internal parallelization of their BLAS backend, the GPRat library~\cite{helmann2025gprat} employs an asynchronous tiled algorithm approach using the HPX runtime system~\cite{kaiser2020hpx}.
This design aims to reduce runtime and memory consumption by utilizing smaller tiles and executing sequential BLAS operations on different tiles in parallel.
The goal of this work is to extend the GPRat library with support for NVIDIA GPUs to accelerate GP predictions.

The main contributions of this work include:
\begin{itemize}
    \item Extension of the GPRat library with a fully GPU-resident GP prediction using CUDA libraries and the task-based runtime system HPX.
    \item Performance evaluation of the tiled Cholesky decomposition for multiple streams with HPX against the cuSOLVER reference.
    \item Performance comparison of the \textit{Cholesky} and \textit{Prediction with Full Covariance} operations between running on the CPU and GPU.
\end{itemize}

The remainder of this work is organized as follows.
\Cref{sec:related-work} provides an overview of related work.
\Cref{sec:software} describes the software stack used in GPRat, including HPX and the specific CPU and GPU backends.
\Cref{sec:methods} details the methods, covering GP regression, tiled algorithms, and the specific GPU implementation.
\Cref{sec:results} presents the performance analysis and CPU versus GPU comparison. 
Finally, in \Cref{sec:conclusion} we conclude and outline future work.

\section{Related Work}
\label{sec:related-work}

GPs are implemented in several Python libraries.
The most popular libraries are GPyTorch~\cite{gardner2018gpytorch} and GPflow~\cite{matthews2017gpflow}.
GPyTorch leverages the cuSOLVER~\cite{cusolver} or MAGMA library~\cite{ahmad2024magma} for linear algebra, utilizing PyTorch's dynamic computational graphs.
It supports Blackbox Matrix-Matrix multiplication (BBMM) and LanczOs Variance Estimates (LOVE)~\cite{pleiss2018love} for scalable inference.
GPflow builds on TensorFlow, benefiting from its scalability and accelerator support.
Recently, GPJax~\cite{pinder2022gpjax} has emerged using the JAX~\cite{bradbury2018jax} framework.
Other libraries like scikit-learn~\cite{pedregosa2011scikit} and GPy~\cite{gpy2014} offer limited or no GPU support.
Their core logic is primarily written in Python, which limits their scalability compared to the C\texttt{++} backends of TensorFlow and PyTorch.

To achieve maximum hardware utilization, parallel programming models are essential.
Asynchronous many-task runtimes (AMTRs) have emerged as powerful tools for tackling irregular problems, such as the Cholesky decomposition. 
A taxonomy of existing AMTRs is given in~\cite{Thoman2018}. Several AMTRs are further compared in a recent survey by Schuchart et. al.~\cite{Schuchart2025}. 
The ExaGeoStat library~\cite{Abdulah2018_starpu_cholesky} uses GPs for geostatistical modeling. Originally based on StarPU~\cite{Augonnet2011_starpu}, ExaGeoStat is also compatible with PaRSEC~\cite{Boscila2013_parsec, Cao2022_parsec_cholesky_gp_cpu} for better scalability.

In this work, we use the HPX library~\cite{kaiser2020hpx}, which is fully compliant with the C\texttt{++} standard API and supports task concatenation via futures.
Applications apart from GPRat~\cite{helmann2025gprat} include HPX-FFT~\cite{Strack2024_hpxfft} and the astrophysics code Octo-Tiger~\cite{Marcello2021}.

\section{Software Stack}
\label{sec:software}

The GPRat library is built upon a robust software stack that combines the performance of C\texttt{++} with the ease-of-use of Python. 
In this section, we present HPX, which we use for asynchronous parallelization, and the BLAS backends for CPU and GPU.

\subsection{HPX}
\label{sec:hpx}
We utilize the HPX~\cite{kaiser2020hpx} runtime system for asynchronous task management.
HPX provides an asynchronous task-based programming model that avoids global synchronization barriers common in fork-join models such as OpenMP~\cite{gabriel2004openmp} or message-based models such as MPI~\cite{mpi40}.
Instead, it relies on fine-grained tasks and lightweight threads.
The core of GPRat's parallelism relies on \texttt{hpx::async} and \texttt{hpx::dataflow}, which schedule tasks based on futurized data dependencies. 
The \texttt{hpx::dataflow} construct allows for implicit dependency management.
We wrap each matrix tile in a \texttt{hpx::shared\_future} object.
To access the results of computations, the \texttt{get()} member function is employed, which synchronizes the execution by blocking only when a specific matrix tile is required.
Consequently, this dependency-driven execution model minimizes idle time and maximizes resource utilization by effectively hiding synchronization latencies through the overlapping of computation and data access. 
HPX can utilize APEX~\cite{huck2022apex}, which allows efficient performance measurements of individual computation steps without synchronization barriers. In this work, we add support for APEX to GPRat.

\subsection{CPU Backends}
For the CPU backend, GPRat relies on highly optimized BLAS and LAPACK implementations.
In this work, we use Intel oneMKL~\cite{intel2024onemkl}, taking advantage of its vectorized math routines. 
Since GPRat manages parallelism via HPX at the tile level, we configure Intel oneMKL to run sequentially to avoid oversubscription and interference between the HPX scheduler and the internal threading of Intel oneMKL.
This ensures that the runtime system has full control over the distribution of work across the available cores.
Note that in recent releases, GPRat also supports OpenBLAS~\cite{openblas} as a more portable alternative to Intel oneMKL.

\subsection{GPU Backends}
\label{sec:cuda_stack}
The GPU-resident implementation proposed in this work uses the CUDA Toolkit.
Specifically, we use the cuBLAS~\cite{cublas} library for standard BLAS operations and the cuSOLVER~\cite{cusolver} library for the Cholesky decomposition on individual tiles.
These operations use a single CUDA stream specified in the function call.
Using different streams allows operations to overlap asynchronously.
Treating CUDA calls as CPU functions (distinguished primarily by a GPU context) allows for seamless integration into the HPX task graph.
This maintains a unified, asynchronous execution model across the host and the device.
However, both CUDA libraries expect column-major matrix ordering.
This differs from the row-major layout standard in Python, C\texttt{++}, and consequently GPRat. 
To account for that difference, we introduce reformulation of the original equations.
By relying on CUDA-specific libraries, our approach limits GPRat to NVIDIA accelerators. However, we plan to support more portable BLAS libraries in future releases.

\section{Methods}
\label{sec:methods}

This section details the methodologies employed in GPRat.
First, we introduce the theoretical foundation of GP regression.
We then describe the tiled algorithm approach, which uses independent tasks to enable asynchronous parallel execution.
Finally, we discuss the GPU implementation.

\subsection{Gaussian Process Regression}
\label{sec:gp}

GPs are a probabilistic regression method used for modeling non-linear system behavior $f: \mathbb{R}^D \to \mathbb{R}, \mathbf x \mapsto y$ using a black-box approach~\cite{görtler2019visual,kocijan2016modelling,kocijan2005nonlinear}.
This approach is particularly valuable when a physical derivation of the system's dynamics is too complex or entirely unavailable.
Instead of relying on explicit physical laws, the model learns the system's behavior directly from observed input-output data.
A feature vector $\mathbf x = [x_1, \dots, x_D] \in \mathbb R^D$ comprises $D \in \mathbb N$ features.
In the context of SI, these features correspond to the $D$ regressors formed by lagged system inputs.
The output $y \in \mathbb R$ represents the corresponding noisy observations.
GPs model this relationship probabilistically, defined as $f(\mathbf x) \sim \text{GP}(m(\mathbf x), k(\mathbf x_i, \mathbf{x}_j))$.
For simplicity, we assume a zero mean function $m(\mathbf x) = 0$.
The covariance function $k(\mathbf x_i, \mathbf{x}_j)$, also known as the kernel, measures the similarity between feature vectors.
In this work, we utilize the squared exponential kernel~\cite{williams2006gaussian}:
\begin{align}
    k(\mathbf x_i, \mathbf{x}_j) := v \cdot \exp \left( -\frac{1}{2\ell} \sum_{d=1}^D (x_{i_d} - x_{j_d})^2 \right) + \delta_{i,j} \sigma^2.
    \label{eq:kernel}
\end{align}
This kernel depends on three hyperparameters: the lengthscale $\ell$, the vertical lengthscale $v$, and the noise variance $\sigma^2$.
While hyperparameters are typically subject to optimization, we utilize fixed default values of $\ell=1, v=1, \sigma^2 = 0.1$ for the experiments presented here for simplicity.

Prediction relies on a training set with $n$ training samples as a feature matrix $\mathbf X \in \mathbb{R}^{n \times D}$ and observations $\mathbf y \in \mathbb R^n$.
To predict $\hat n$ new observations $\mathbf{\hat y} \in \mathbb{R}^{\hat n}$ for a test feature matrix $\mathbf{\hat X} \in \mathbb R^{\hat n \times D}$, we have to assemble multiple covariance matrices where each entry $(i,j)$ corresponds to the kernel evaluation $k(\mathbf a_i, \mathbf b_j)$ for feature vectors $\mathbf a_i$ and $\mathbf b_j$ taken from $\mathbf X$ or $\mathbf{\hat X}$.
We define the training covariance matrix $\mathbf K := \mathbf K_{\mathbf X, \mathbf X} \in \mathbb R^{n \times n}$, the cross-covariance matrix $\smash{\mathbf K_{\mathbf{\hat X}, \mathbf X}} \in \mathbb R^{\hat n \times n}$, and the prior test covariance matrix $\mathbf K_{\mathbf{\hat X}, \mathbf{\hat X}} \in \mathbb R^{\hat n \times \hat n}$.
The predictive mean is calculated as:
\begin{equation}
    \mathbf{\hat y} := \mathbf K_{\mathbf{\hat X},\mathbf X} \cdot \mathbf K^{-1} \cdot \mathbf y.
    \label{eq:gp_mean}
\end{equation}
Direct inversion of $\mathbf K$ is avoided to ensure numerical stability.
Instead, we exploit the symmetric, positive semi-definite property of $\mathbf K$ to apply the Cholesky decomposition $\mathbf K = \mathbf L \cdot \mathbf L^\top$, where $\mathbf L$ is a lower triangular matrix.
The prediction is then computed efficiently by solving $\mathbf L \cdot \boldsymbol \beta = \mathbf y$ (forward substitution), followed by $\mathbf L^\top \cdot \boldsymbol \alpha = \boldsymbol \beta$ (backward substitution), and finally $\mathbf{\hat y} = \mathbf K_{\mathbf{\hat X}, \mathbf X} \cdot \boldsymbol \alpha$.

Simultaneously, GPs provide uncertainty estimates via the posterior covariance matrix $\mathbf\Sigma_{\mathbf{\hat X}}$.
While the matrix $\mathbf K$ represents prior covariances based on kernel evaluations, $\mathbf\Sigma_{\mathbf{\hat X}}$ represents the conditional covariance after observing the training data.
The diagonal elements of $\mathbf \Sigma_{\mathbf{\hat X}}$ represent the predictive variances, derived from:
\begin{equation}
    \mathbf \Sigma_{\mathbf{\hat X}} := \mathbf K_{\mathbf{\hat X}, \mathbf{\hat X}} - \mathbf K_{\mathbf{\hat X}, \mathbf X} \cdot \mathbf K^{-1} \cdot \mathbf K_{\mathbf X, \mathbf{\hat X}}.
    \label{eq:posterior-covariance}
\end{equation}
Computationally, this is handled by solving $\mathbf L \cdot \mathbf V = \smash{\mathbf K_{\mathbf X, \mathbf{\hat X}}}$ for $\mathbf V$, computing the temporary matrix $\mathbf W := \mathbf V^\top \cdot \mathbf V$, and finally obtaining $\mathbf \Sigma_{\mathbf{\hat X}} = \mathbf K_{\mathbf{\hat X}, \mathbf{\hat X}} - \mathbf W$.
For large test datasets, the latter steps are increasingly computationally expensive compared to the Cholesky decomposition.

\subsection{Tiled Algorithms}
\label{sec:tiled-algorithms}

To handle the computational complexity of GPs, specifically the cubic complexity of the Cholesky decomposition, we employ tiled algorithms.
These algorithms parallelize GP operations by dividing the computation into smaller tasks distributed across multiple hardware cores.
This asynchronous, task-based approach has been shown to offer superior scalability and parallel efficiency compared to standard reference implementations~\cite{strack2023scalability}.
Our primary focus is the tiled Cholesky decomposition of the symmetric, positive semi-definite covariance matrix $\mathbf K \in \mathbb R^{n \times n}$.
This tiled approach offers significant memory advantages, requiring only \SI{50}{\percent} to \SI{75}{\percent} of the memory compared to storing the entire symmetric matrix.

The matrix $\mathbf K$ is split into $M \times M$ tiles $\mathbf K_{I,J} \in \mathbb R^{m \times m}$, where $m = n/M$.
The tiled algorithm utilizes a \emph{right-looking} strategy~\cite{buttari2009right}. There also exist \emph{left-} and \emph{top-looking} variants~\cite{dorris2016right2}.
The tiled algorithm relies on four standard BLAS operations optimized for memory access~\cite{kowarschik2003cache}:
\begin{enumerate}
    \item \textcolor{potrf}{\textbf{POTRF}}: Cholesky decomposition on diagonal tile ($ \mathbf L_{J,J} \leftarrow \text{Chol}(\mathbf K_{J,J})$).
    \item \textcolor{trsm}{\textbf{TRSM}}: Solve triangular system on current column ($\mathbf L_{I,J} \leftarrow \mathbf L_{J,J}^{-1} \mathbf K_{I,J}$).
    \item \textcolor{syrk}{\textbf{SYRK}}: Update diagonal tiles ($\mathbf K_{I,I} \leftarrow \mathbf K_{I,I} - \mathbf L_{I,J} \mathbf L_{I,J}^\top$).
    \item \textcolor{gemm}{\textbf{GEMM}}: Update off-diagonal tiles ($\mathbf K_{I,K} \leftarrow \mathbf K_{I,K} - \mathbf L_{I,J} \mathbf L_{K,J}^\top$).
\end{enumerate}

\Cref{fig:tiled-algo,fig:tiled-vis} present the algorithmic structure alongside a visualization of the matrix state during the first iteration ($J=0$).
In this step, POTRF is executed on the top-left diagonal matrix tile.
Subsequently, TRSM is applied to all matrix tiles in the column below.
The resulting tiles are then used to update the trailing sub-matrix via SYRK and GEMM, effectively preparing the matrix for the next iteration ($J=1$).

\begin{figure}[t]
    \centering
    \begin{minipage}[c]{0.48\textwidth}
        \begin{algorithmic}[1]
        \footnotesize
        \For{$J \gets 0$ \textbf{to} $M - 1$}
            \State \textcolor{potrf}{\textbf{POTRF}}($\mathbf K_{J,J}$)
            \For{$I \gets J + 1$ \textbf{to} $M - 1$}
                \State \textcolor{trsm}{\textbf{TRSM}}($\mathbf K_{J,J}, \mathbf K_{I,J}$)
            \EndFor
            \For{$I \gets J + 1$ \textbf{to} $M - 1$}
                \State \textcolor{syrk}{\textbf{SYRK}}($\mathbf K_{I,J}, \mathbf K_{I,I}$)
                \For{$K \gets J + 1$ \textbf{to} $I - 1$}
                    \State \textcolor{gemm}{\textbf{GEMM}}($\mathbf K_{I,J}, \mathbf K_{K,J}, \mathbf K_{I,K}$)
                \EndFor
            \EndFor
        \EndFor
        \end{algorithmic}
        \caption{Tiled Cholesky decomposition of $\mathbf K$.}
        \label{fig:tiled-algo}
    \end{minipage}
    \hfill
    \begin{minipage}[c]{0.48\textwidth}
        \centering
        \begin{tikzpicture}[scale=0.65]
            \fill[utile] (1, 0) rectangle ++(1, -1);
            \fill[utile] (2, 0) rectangle ++(1, -1);
            \fill[utile] (3, 0) rectangle ++(1, -1);
            \fill[utile] (4, 0) rectangle ++(1, -1);

            \fill[utile] (2, -1) rectangle ++(1, -1);
            \fill[utile] (3, -1) rectangle ++(1, -1);
            \fill[utile] (4, -1) rectangle ++(1, -1);

            \fill[utile] (3, -2) rectangle ++(1, -1);
            \fill[utile] (4, -2) rectangle ++(1, -1);

            \fill[utile] (4, -3) rectangle ++(1, -1);

            \fill[potrf] (0, 0) rectangle ++(1, -1);

            \fill[trsm] (0, -1) rectangle ++(1, -1);
            \fill[trsm] (0, -2) rectangle ++(1, -1);
            \fill[trsm] (0, -3) rectangle ++(1, -1);
            \fill[trsm] (0, -4) rectangle ++(1, -1);

            \fill[syrk] (1, -1) rectangle ++(1, -1);
            \fill[syrk] (2, -2) rectangle ++(1, -1);
            \fill[syrk] (3, -3) rectangle ++(1, -1);
            \fill[syrk] (4, -4) rectangle ++(1, -1);

            \fill[gemm] (1, -2) rectangle ++(1, -1);

            \fill[gemm] (1, -3) rectangle ++(1, -1);
            \fill[gemm] (2, -3) rectangle ++(1, -1);

            \fill[gemm] (1, -4) rectangle ++(1, -1);
            \fill[gemm] (2, -4) rectangle ++(1, -1);
            \fill[gemm] (3, -4) rectangle ++(1, -1);

            \draw[step=1cm, white, line width=0.5mm] (0,0) grid (5,-5);

            \node[black] at (0.5, -0.5) {\textbf{P}};

            \node[black] at (0.5, -1.5) {\textbf{T}};
            \node[black] at (0.5, -2.5) {\textbf{T}};
            \node[black] at (0.5, -3.5) {\textbf{T}};
            \node[black] at (0.5, -4.5) {\textbf{T}};

            \node[black] at (1.5, -1.5) {\textbf{S}};
            \node[black] at (2.5, -2.5) {\textbf{S}};
            \node[black] at (3.5, -3.5) {\textbf{S}};
            \node[black] at (4.5, -4.5) {\textbf{S}};

            \node[white] at (1.5, -2.5) {\textbf{G}};

            \node[white] at (1.5, -3.5) {\textbf{G}};
            \node[white] at (2.5, -3.5) {\textbf{G}};

            \node[white] at (1.5, -4.5) {\textbf{G}};
            \node[white] at (2.5, -4.5) {\textbf{G}};
            \node[white] at (3.5, -4.5) {\textbf{G}};

            \foreach \i in {0,...,4} {
                \node[black, font=\footnotesize] at (-0.3, -\i-0.5) {\i};
                \node[black, font=\footnotesize] at (\i+0.5, 0.3) {\i};
            }
        \end{tikzpicture}
        \caption{Example for matrix $\mathbf K$ split into $5 \times 5$ tiles, colored by tasks in the first iteration ($J=0$): \textcolor{potrf}{\textbf{POTRF}}, \textcolor{trsm}{\textbf{TRSM}}, \textcolor{syrk}{\textbf{SYRK}}, and \textcolor{gemm}{\textbf{GEMM}}.}
        \label{fig:tiled-vis}
    \end{minipage}
\end{figure}

\subsection{GPU Implementation}
\label{sec:gpu-implementation}

In this work, we accelerate the tiled algorithms by leveraging NVIDIA GPUs.
We utilize the cuBLAS~\cite{cublas} and cuSOLVER~\cite{cusolver} libraries for standard linear algebra routines, while custom CUDA kernels handle GP-related tasks such as the asynchronous assembly of the covariance matrix.
We prioritize minimizing data transfer overhead; therefore, we employ manual memory management rather than Unified Memory, which has been shown to incur performance penalties~\cite{nadal2016unified}.
Both training and test datasets are explicitly copied to the device, all computations are performed entirely in device memory, and only the final results are retrieved to the host.
Consequently, the tiled matrices in the GPU implementation are represented as \texttt{std::vector<hpx::shared\_future<double*{>}{>}}, where each future holds a pointer to the specific matrix tile in device memory.

Regarding stream management, HPX offers executors for CUDA and cuBLAS kernels but currently lacks support for cuSOLVER.
To maintain consistency across all library calls and ensure precise control over execution, we rely on CUDA streams and handles directly.
GPRat initializes a user-configurable pool of CUDA streams and handles.
To manage the large number of independent tasks generated by the tiled algorithms, we employ a simple round-robin scheduling strategy to assign streams to tasks.
This strategy distributes operations across the available streams to facilitate concurrent execution without the substantial overhead of creating and destroying streams for each task.
Stream synchronization is handled via \texttt{cudaStreamSynchronize} directly within the HPX tasks calling cuBLAS and cuSOLVE kernels, allowing the HPX runtime to manage task dependencies via \texttt{hpx::dataflow} and ensure asynchronous execution.

Another significant challenge in integrating standard CUDA libraries with GPRat is the different memory layout.
GPRat, designed for compatibility with Python and C\texttt{++}, utilizes row-major storage with 0-based indexing.
In contrast, cuBLAS and cuSOLVER assume column-major storage with 1-based indexing.
Explicitly transposing matrices in memory introduces substantial overhead.
Instead, we exploit the algebraic property that a matrix $\mathbf A$ stored in row-major order is identical in memory to its transpose $\mathbf A^\top$ stored in column-major order.
Consequently, we adapt the BLAS calls by logically transposing the matrix equations.

For example, for the TRSM operation, we transpose the defining equation to map it to column-major routines:
\begin{align}
    \mathbf K_{J,J}^\top \cdot \mathbf L_{I,J} &= \mathbf K_{I,J} \\
    \iff (\mathbf K_{J,J}^\top \cdot \mathbf L_{I,J})^\top &= \mathbf K_{I,J}^\top \\
    \iff \mathbf L_{I,J}^\top \cdot (\mathbf K_{J,J}^\top)^\top &= \mathbf K_{I,J}^\top.
\end{align}
In this transformed context, we ensure that every matrix appears transposed at least once, allowing it to be interpreted as a column-major matrix in standard BLAS routines.
For instance, $(\mathbf K_{J,J}^\top)^\top$ represents the row-major stored $\mathbf K_{J,J}$ interpreted as a column-major matrix that must be transposed.
Therefore, we invoke \texttt{cublasDtrsm} with the \texttt{CUBLAS\_OP\_T} flag.
Additionally, the multiplication ordering changes, requiring the side parameter to be switched.
Similar transformations are applied to SYRK and GEMM operations to ensure correct mathematical results without explicit data rearrangement.
For functions involving symmetric matrices where only one triangular part is referenced (e.g., POTRF), we utilize the fact that the lower triangular part of a row-major matrix corresponds to the upper triangular part of a column-major matrix, passing \texttt{CUBLAS\_FILL\_MODE\_UPPER} accordingly.

\section{Results}
\label{sec:results}

In this section, we present the results of our Cholesky decomposition analysis and performance comparison between CPU and GPU.
We evaluated the performance of the GPRat library on a single node equipped with a dual-socket AMD EPYC 9274F CPU with a total of $2 \times 24$ cores and theoretical peak performance of $\num{3.1}$~TFLOPS and a TDP of 640\si{\watt}. The node also features a NVIDIA A30 GPU (\SI{24}{\giga\byte}) with a theoretical peak performance of $\num{5.2}$~TFLOPS and a TDP of 165\si{\watt}.
The software stack includes GPRat $v0.3.1$, HPX $1.10.0$, CUDA $12.0.1$, and Intel oneMKL $2024.2.2$.
Runtimes were averaged over $50$ runs for runs shorter than $\SI{1e{-2}}{\second}$ and ten runs for longer runs. Error bars represent the \SI{95}{\percent} confidence interval.

\subsection{Cholesky Decomposition}

\begin{figure}[t]
    \centering
    \begin{minipage}[t]{.47\textwidth}
        \centering
        \includegraphics[width=\linewidth]{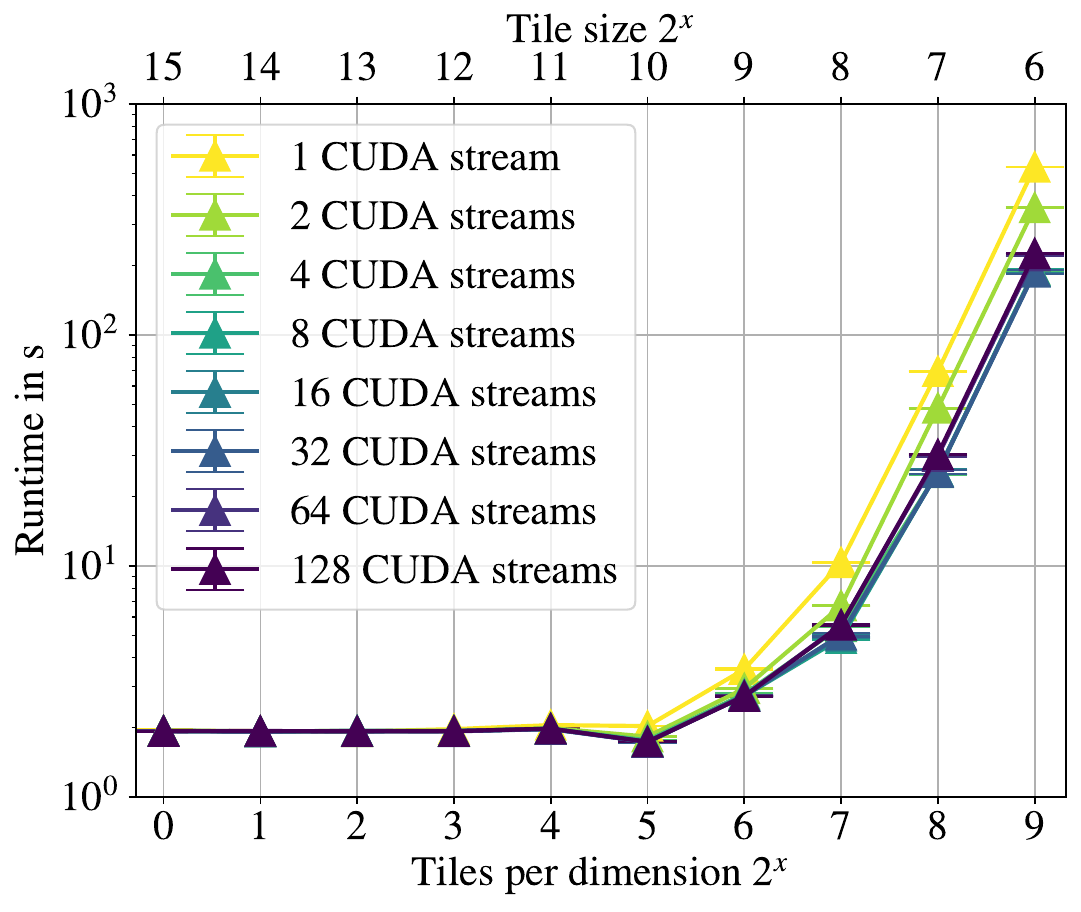}
        \caption{Runtime of Cholesky on GPU for a problem size of $n=\num{32768}$ with varying CUDA streams and tiles.}
        \label{fig:tile-scaling-cholesky-gpu-streams}
    \end{minipage}\hspace{.05\textwidth}
    \begin{minipage}[t]{.47\textwidth}
        \centering
        \includegraphics[width=\linewidth]{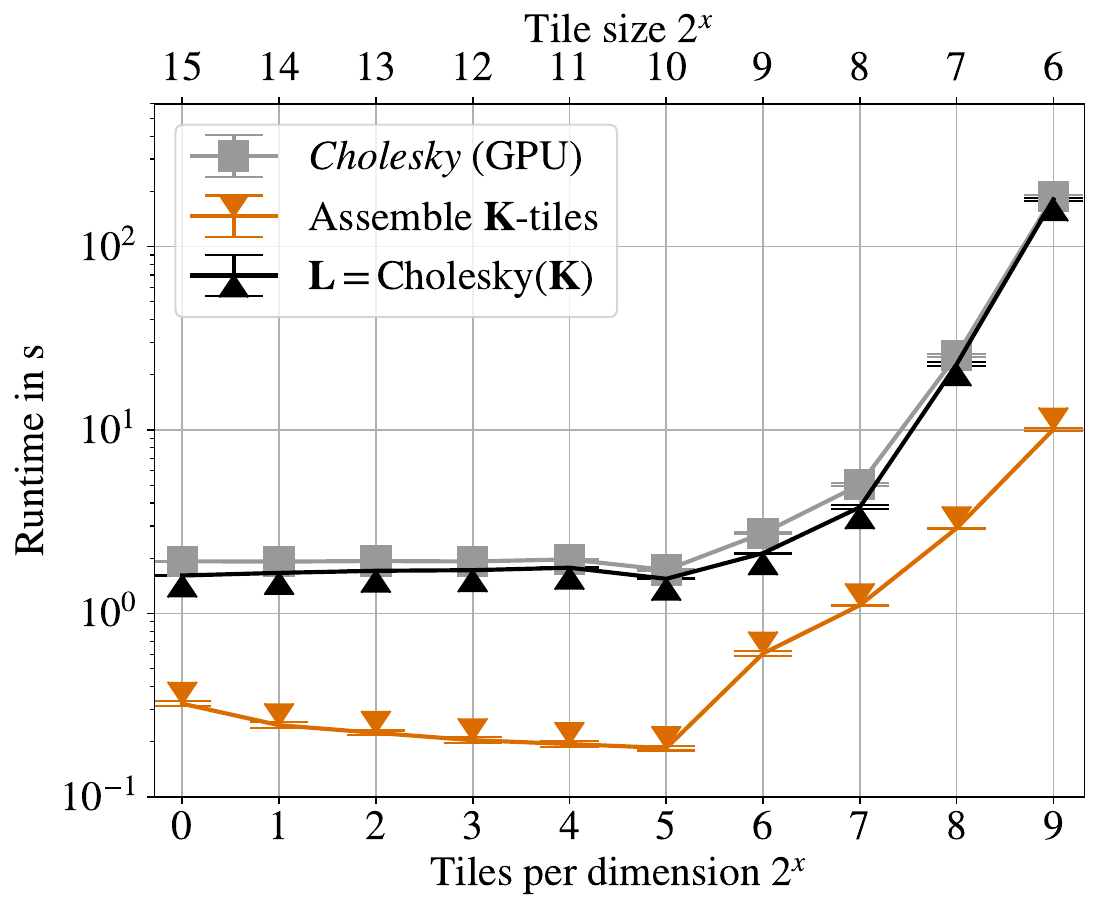}
        \caption{Breakdown of Cholesky runtime steps on GPU for a problem size of $n=\num{32768}$ and $32$ streams with varying tiles.}
        \label{fig:tile-scaling-cholesky-gpu-steps}
    \end{minipage}
\end{figure}

We analyzed the runtime of the Cholesky decomposition both on CPU and GPU.
Using tiled algorithms yields the additional benefit of reduced memory consumption, as only non-empty tiles for the lower triangular factor need to be stored.
In terms of CPU performance, on our 48-core test system, we observed a speedup of approximately $43.7$ compared to sequential execution.
Furthermore, the assembly of the covariance matrix has a quadratic time complexity $\mathcal O(n^2)$ and is therefore significantly faster than the Cholesky decomposition with cubic time complexity $\mathcal O(n^3)$.
For the GPU implementation, we investigated the scalability with respect to the number of tiles and CUDA streams.
\Cref{fig:tile-scaling-cholesky-gpu-streams} shows that using 32 CUDA streams with 32 tiles per dimension yields the optimal performance for $n=\num{32768}$.
This configuration achieves a runtime reduction of approximately \SI{11}{\percent} compared to using a single tile.
Using a single tile is equivalent to a pure cuSOLVER call, meaning our tiled approach with $32$ streams outperforms using the cuSOLVER library in a direct comparison.
The overhead of managing more streams or tiles outweighs the benefits of increased parallelism beyond this point.
Decomposing the total runtime into its components (see \Cref{fig:tile-scaling-cholesky-gpu-steps}) confirms that the Cholesky factorization itself, although the dominant factor, benefits from parallelization when a specific tiling of the covariance matrix is used.

\begin{figure}[t]
    \centering
    \includegraphics[width=\textwidth]{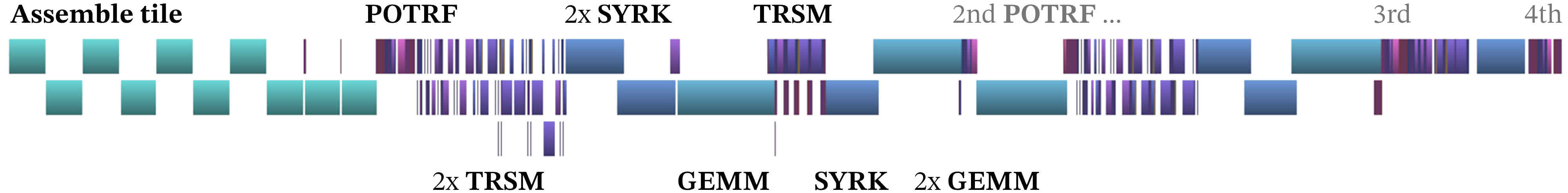}
    \caption{Cholesky in NVIDIA Visual Profiler for $n = \num{4096}$ training samples, four tiles, and four streams.}
    \label{fig:cholesky-1050}
\end{figure}

Additionally, we visualize the Cholesky decomposition using the NVIDIA Visual Profiler on a GTX 1050 GPU (see \Cref{fig:cholesky-1050}) to show the scheduling and execution of the tiled operations on the GPU.
The timeline flows from left to right and contains rectangles that represent CUDA kernel calls.
Three rows are used to draw overlapping kernel calls, revealing limited concurrency, with overlap primarily at the start and end of kernel execution.
Note the asynchronous scheduling of TRSM and trailing sub-matrix updates.
The cuSOLVER POTRF itself applies a parallel algorithm based on tiled operations, executing multiple CUDA kernels sequentially within a single CUDA stream, where each kernel parallelizes the computation of matrix entries across GPU threads. 
This inherent parallelism within the kernels typically saturates GPU resources, resulting in smaller performance improvements from the HPX-based tiled algorithm compared to the CPU variant, which uses a sequential backend.
The assembly kernels do not overlap. 
However, in Figure \ref{fig:tile-scaling-cholesky-gpu-steps} we observe a runtime reduction for the covariance matrix assembly.
This is caused by the tiled data structure of the covariance matrix, for which fewer entries have to be computed with increasing tiles.

\subsection{CPU and GPU Comparison}

\begin{figure}[t]
    \centering
    \begin{minipage}[t]{.47\textwidth}
        \centering
        \includegraphics[width=\linewidth]{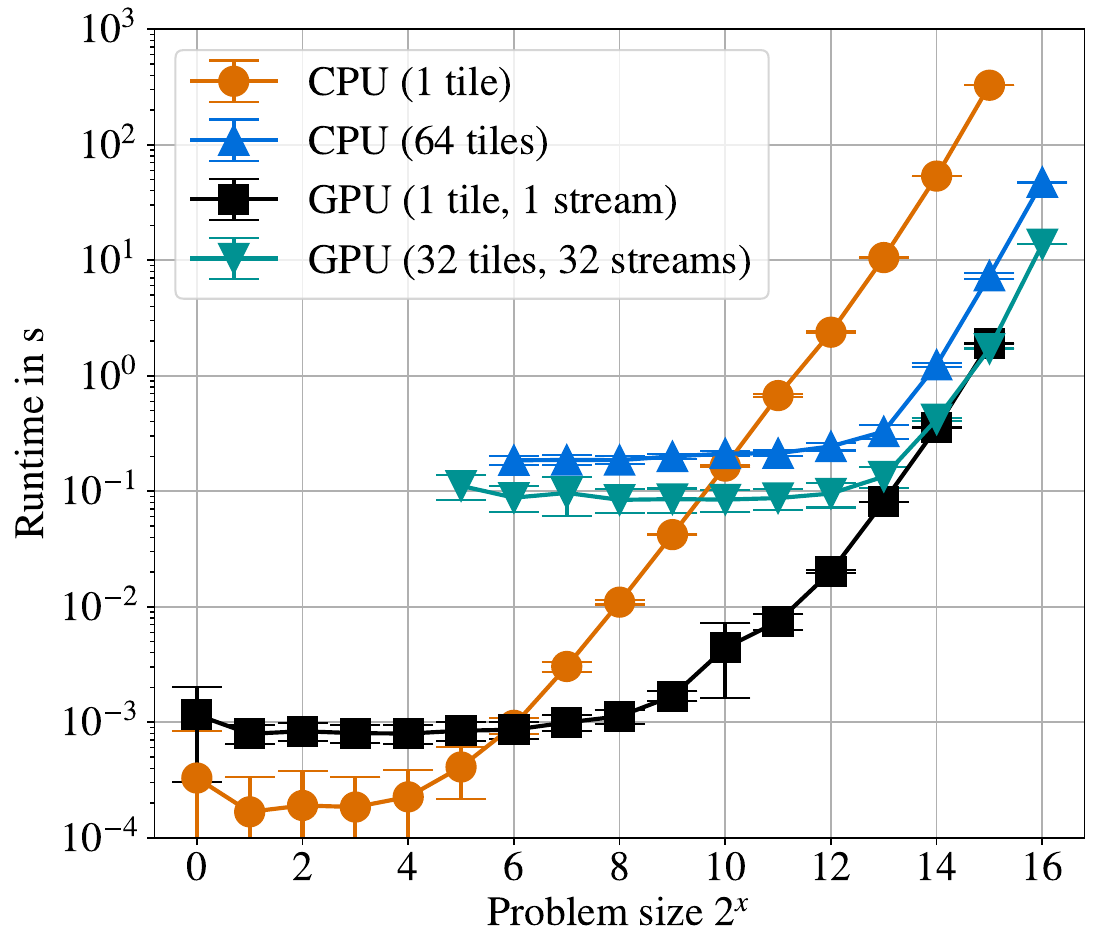}
        \caption{\emph{Cholesky} problem size scaling comparison between CPU and GPU.}
        \label{fig:problem-size-cholesky-cpu-gpu}
    \end{minipage}\hspace{.05\textwidth}
    \begin{minipage}[t]{.47\textwidth}
        \centering
        \includegraphics[width=\linewidth]{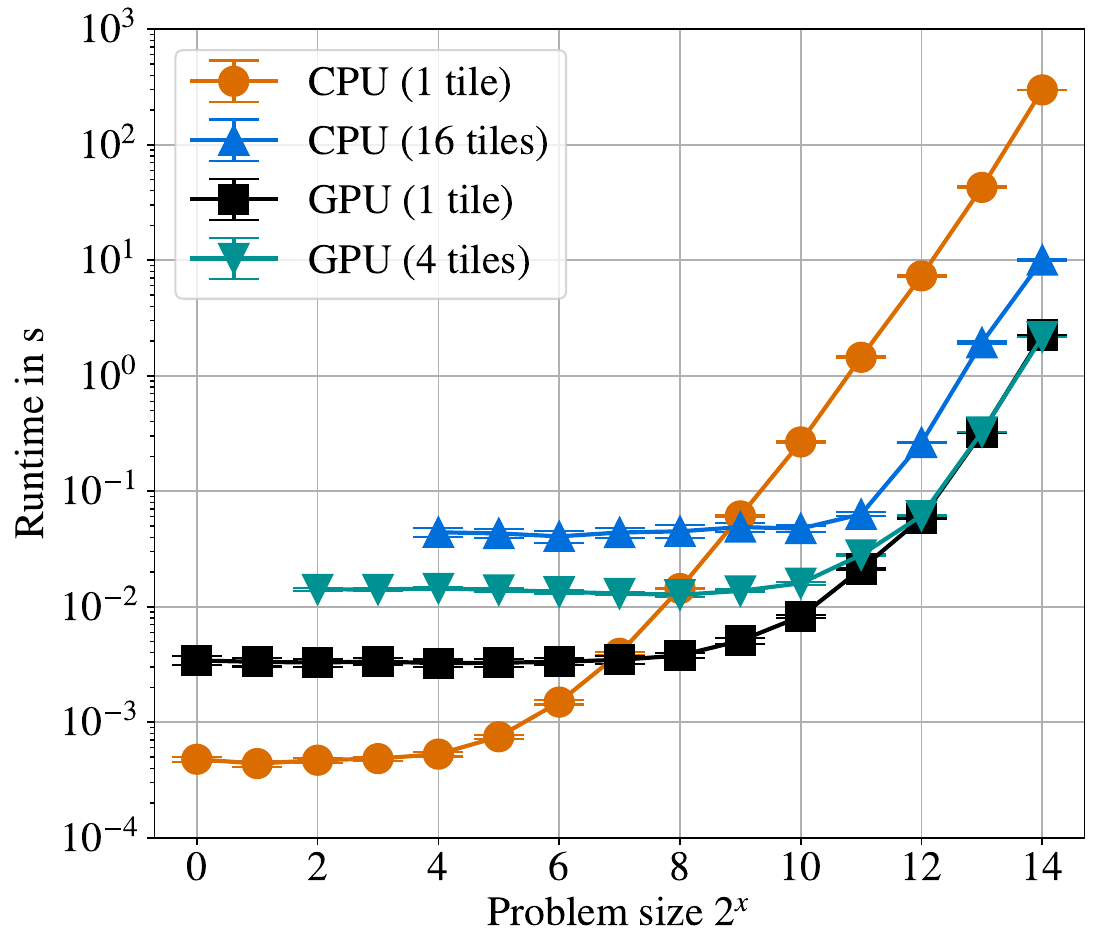}
        \caption{\emph{Predict Full Covariance} problem size scaling comparison between CPU and GPU.}
        \label{fig:problem-size-predict-fullcov-cpu-gpu}
    \end{minipage}
\end{figure}

Comparing CPU and GPU performance for \emph{Cholesky} (see \Cref{fig:problem-size-cholesky-cpu-gpu}), the GPU implementation outperforms the CPU for problem sizes $n \ge 128$.
For smaller problem sizes, the sequential execution on a single tile is faster due to task management overheads in the HPX runtime.
For $n=\num{32768}$, the GPU implementation (with $32$ tiles per dimension) achieves a speedup of approximately $\num{4.3}$ over the CPU implementation (with $64$ tiles per dimension).
For the \emph{Predict with Full Covariance Matrix} operation (see \Cref{fig:problem-size-predict-fullcov-cpu-gpu}), where the number of test samples $\hat{n}$ corresponds to the number of training samples $n$, the GPU implementation achieves a speedup of $\num{4.6}$ over the CPU version at $n=\num{16384}$. 
Although the theoretical peak FLOPS of our test system would suggest a speedup of roughly $\num{1.7}$, in practice, the GPU implementation achieves a significantly higher speedup.
This is caused by CPU strong scaling limitations on our system, related to the dual-socket configuration and chiplet design of contemporary AMD CPUs~\cite{helmann2025gprat}. 
Our results highlight the superiority of modern accelerators for compute-heavy workloads.
In our benchmarks, the GPU delivers over four times the performance while consuming only one quarter of the energy.

\section{Conclusion and Outlook}
\label{sec:conclusion}

In this work, we extended the GPRat library with a fully GPU-resident GP prediction combining CUDA and HPX.
The implementation of tiled algorithms for the Cholesky decomposition and prediction allows efficient distribution of tasks across the available hardware.
Our results show that the GPU implementation achieves significant speedups over the CPU implementation for datasets larger than ($n \ge 128$) samples. We achieve speedups of up to $\num{4.3}$ for Cholesky decomposition itself and speedups of up to $\num{4.6}$ for the GP prediction and uncertainty estimation.
Through the use of multiple CUDA streams, our approach can beat cuSOLVER by up to \SI{11}{\percent} for the right configuration, even though we add additional task creation and scheduling overhead with the HPX runtime.

Future work will focus on optimizations to further improve the utilization of the GPU.
Since distributed computing is supported by the underlying HPX runtime, we aim to extend the library to distributed multi-GPU environments to overcome single-node memory limitations.
Additionally, we plan to implement heterogeneous computing to leverage both CPU and GPU resources simultaneously, ensuring maximum hardware utilization.
Finally, we plan to introduce mixed-precision calculations to take advantage of modern hardware capabilities.

\begin{credits}

\subsection*{Supplementary Materials}
The GPRat library is publicly available as open-source software under the MIT License on \href{https://github.com/SC-SGS/GPRat}{GitHub}\footnote{\url{https://github.com/SC-SGS/GPRat} Accessed: 2026-01-10}. For result reproduction, use GPRat release v$0.3.1$, which includes NVIDIA GPU support as well as special compilation flags to measure the runtime of individual computation steps using APEX. Datasets of arbitrary size can be generated with GPRat's mass-spring-damper simulator. A reference dataset is available on \href{https://github.com/SC-SGS/GPRat}{DaRUS}\footnote{\url{https://doi.org/10.18419/DARUS-4743} Accessed: 2026-01-10}.

\subsection*{AI Use Disclosure}

Generative AI tools, including Grammarly~\cite{grammarly}, DeepL~\cite{deepl}, Gemini~\cite{gemini}, and ChatGPT~\cite{chatgpt}, were employed to enhance the clarity, grammar, and overall coherence of the manuscript. All technical content, data analyses, and research findings were conceived and developed independently by the authors. AI-assisted outputs were carefully reviewed, verified, and edited by the authors to ensure factual accuracy, interpretive rigor, and scholarly integrity. The final manuscript reflects the authors’ original intellectual contributions and analytical work.

\end{credits}

\bibliographystyle{splncs04}
\bibliography{main}

@inproceedings{strack2023scalability,
  address   = {Cham},
  author    = {Strack, Alexander and Pfl{\"u}ger, Dirk},
  booktitle = {Asynchronous Many-Task Systems and Applications},
  pages     = {52--64},
  publisher = {Springer Nature Switzerland},
  title     = {{Scalability of~Gaussian Processes Using Asynchronous Tasks: A Comparison Between HPX and~PETSc}},
  year      = {2023}
}

@manual{intel2024onemkl,
  author       = {{Intel Corporation}},
  organization = {Intel Corporation},
  title        = {{Intel\textregistered{} oneAPI Math Kernel Library Version 2024.2.2}},
  year         = {2024}
}

@inproceedings{gardner2018gpytorch,
  author    = {Gardner, Jacob R and Pleiss, Geoff and Bindel, David and others},
  booktitle = {Advances in Neural Information Processing Systems},
  title     = {{GPyTorch: Blackbox Matrix-Matrix Gaussian Process Inference with GPU Acceleration}},
  year      = {2018}
}

@article{pleiss2018love,
  author       = {Geoff Pleiss and
                  Jacob R. Gardner and
                  Kilian Q. Weinberger and
                  others},
  title        = {Constant-Time Predictive Distributions for Gaussian Processes},
  journal      = {CoRR},
  volume       = {abs/1803.06058},
  year         = {2018},
}

@article{matthews2017gpflow,
  author  = {Matthews, Alexander G. de G. and {van der Wilk}, Mark and Nickson, Tom and others},
  journal = {Journal of Machine Learning Research},
  month   = {apr},
  number  = {40},
  pages   = {1--6},
  title   = {{{GP}flow: A {G}aussian process library using {T}ensor{F}low}},
  volume  = {18},
  year    = {2017}
}

@manual{cublas,
  author       = {{NVIDIA Corporation}},
  howpublished = {\url{https://docs.nvidia.com/cuda/cublas/index.html}},
  organization = {NVIDIA Corporation},
  title        = {{cuBLAS} {Library} {User} {Guide}},
  note         = {Accessed: 2026-01-10},
  year         = {2026}
}

@manual{cusolver,
  author       = {{NVIDIA Corporation}},
  howpublished         = {\url{https://docs.nvidia.com/cuda/cusolver/index.html}},
  organization = {NVIDIA Corporation},
  title        = {{cuSOLVER} {Library} {User} {Guide}},
  note         = {Accessed: 2026-01-10},
  year         = {2026}
}

@article{kaiser2020hpx,
  author    = {Hartmut Kaiser and Patrick Diehl and Adrian S. Lemoine and others},
  journal   = {Journal of Open Source Software},
  number    = {53},
  pages     = {2352},
  publisher = {The Open Journal},
  title     = {{HPX - The C++ Standard Library for Parallelism and Concurrency}},
  volume    = {5},
  year      = {2020}
}

@inproceedings{huck2022apex,
  author    = {Huck, Kevin A.},
  booktitle = {
               2022 IEEE/ACM 7th International Workshop on Extreme Scale Programming
               Models and Middleware (ESPM2)},
  pages     = {20--29},
  title     = {{
               Broad Performance Measurement Support for Asynchronous Multi-Tasking
               with APEX}},
  year      = {2022}
}

@article{pedregosa2011scikit,
  author     = {
                Pedregosa, Fabian and Varoquaux, Ga\"{e}l and Gramfort, Alexandre and
                others},
  issue_date = {2/1/2011},
  journal    = {J. Mach. Learn. Res.},
  month      = nov,
  number     = {null},
  numpages   = {6},
  pages      = {2825–2830},
  publisher  = {JMLR.org},
  title      = {{Scikit-learn: Machine Learning in Python}},
  volume     = {12},
  year       = {2011}
}

@inproceedings{berkenkamp2015control,
  author    = {Berkenkamp, Felix and Schoellig, Angela P.},
  booktitle = {2015 European Control Conference (ECC)},
  pages     = {2496--2501},
  title     = {{Safe and robust learning control with Gaussian processes}},
  year      = {2015}
}

@inproceedings{damianou2013deep,
  address   = {Scottsdale, Arizona, USA},
  author    = {Damianou, Andreas and Lawrence, Neil D.},
  booktitle = {
               Proceedings of the Sixteenth International Conference on Artificial
               Intelligence and Statistics},
  month     = {29 Apr--01 May},
  pages     = {207--215},
  publisher = {PMLR},
  title     = {Deep {G}aussian Processes},
  volume    = {31},
  year      = {2013}
}

@article{liu2020bigdata,
  author   = {Liu, Haitao and Ong, Yew-Soon and Shen, Xiaobo and others},
  journal  = {IEEE Transactions on Neural Networks and Learning Systems},
  number   = {11},
  pages    = {4405--4423},
  title    = {{When Gaussian Process Meets Big Data: A Review of Scalable GPs}},
  volume   = {31},
  year     = {2020}
}

@book{kocijan2016modelling,
  author    = {Kocijan, Ju{\v{s}}},
  publisher = {Springer},
  title     = {{Modelling and control of dynamic systems using Gaussian process models}},
  year      = {2016}
}

@incollection{kowarschik2003cache,
  address   = {Berlin, Heidelberg},
  author    = {Kowarschik, Markus and Wei{\ss}, Christian},
  booktitle = {Algorithms for Memory Hierarchies: Advanced Lectures},
  pages     = {213--232},
  publisher = {Springer Berlin Heidelberg},
  title     = {
               {An Overview of Cache Optimization Techniques and Cache-Aware Numerical
               Algorithms}},
  year      = {2003}
}

@article{nadal2016unified,
  author   = {Nadal-Serrano, Jose M. and Lopez-Vallejo, Marisa},
  journal  = {IEEE Transactions on Parallel and Distributed Systems},
  number   = {6},
  pages    = {1579--1588},
  title    = {
              A Performance Study of CUDA UVM versus Manual Optimizations in a
              Real-World Setup: Application to a Monte Carlo Wave-Particle
              Event-Based Interaction Model},
  volume   = {27},
  year     = {2016}
}

@article{ljung2010si,
  author   = {Lennart Ljung},
  journal  = {Annual Reviews in Control},
  number   = {1},
  pages    = {1--12},
  title    = {Perspectives on system identification},
  volume   = {34},
  year     = {2010}
}

@inbook{kocijan2005nonlinear,
  address   = {Berlin, Heidelberg},
  author    = {Kocijan, Ju{\v{s}} and Murray-Smith, Roderick},
  booktitle = {
               Switching and Learning in Feedback Systems: European Summer School on
               Multi-Agent Control, Maynooth, Ireland, September 8-10, 2003, Revised
               Lectures and Selected Papers},
  pages     = {185--200},
  publisher = {Springer Berlin Heidelberg},
  title     = {{Nonlinear Predictive Control with a Gaussian Process Model}},
  year      = {2005}
}

@article{buttari2009right,
  author   = {Alfredo Buttari and Julien Langou and Jakub Kurzak and others},
  journal  = {Parallel Computing},
  number   = {1},
  pages    = {38--53},
  title    = {
              A class of parallel tiled linear algebra algorithms for multicore
              architectures},
  volume   = {35},
  year     = {2009}
}

@inproceedings{dorris2016right2,
  address   = {Cham},
  author    = {
               Dorris, Joseph and Kurzak, Jakub and Luszczek, Piotr and others},
  booktitle = {High Performance Computing},
  pages     = {544--562},
  publisher = {Springer International Publishing},
  title     = {{Task-Based Cholesky Decomposition on Knights Corner Using OpenMP}},
  year      = {2016}
}

@article{görtler2019visual,
  author  = {G\"{o}rtler, Jochen and Kehlbeck, Rebecca and Deussen, Oliver},
  journal = {Distill},
  title   = {{A Visual Exploration of Gaussian Processes}},
  year    = {2019}
}

@book{williams2006gaussian,
  author    = {Williams, Christopher KI and Rasmussen, Carl Edward},
  publisher = {MIT press Cambridge, MA},
  title     = {Gaussian processes for machine learning},
  volume    = {2},
  year      = {2006}
}

@article{pinder2022gpjax,
  author   = {{Pinder}, Thomas and {Dodd}, Daniel},
  journal  = {The Journal of Open Source Software},
  month    = jul,
  number   = {75},
  pages    = {4455},
  title    = {{GPJax: A Gaussian Process Framework in JAX}},
  volume   = {7},
  year     = {2022}
}

@misc{gpy2014,
  author       = {{
Sheffield Machine Learning}},
  howpublished = {\url{http://github.com/SheffieldML/GPy}},
  title        = {{GPy}: A Gaussian process framework in python},
  year         = {2012},
  note         = {Accessed: 2026-01-10}
}

@misc{bradbury2018jax,
  author = {James Bradbury and Roy Frostig and Peter Hawkins and others},
  title = {{JAX}: composable transformations of {P}ython+{N}um{P}y programs},
  url = {http://github.com/jax-ml/jax},
  version = {0.3.13},
  year = {2018},
  note         = {Accessed: 2026-01-10}
}

@article{ahmad2024magma,
  author   = {
              Ahmad Abdelfattah and Natalie Beams and Robert Carson and others},
  journal  = {Int. J. High Perform. Comput. Appl.},
  number   = {5},
  pages    = {468--490},
  title    = {
              {MAGMA: Enabling exascale performance with accelerated BLAS and LAPACK
              for diverse GPU architectures}},
  volume   = {38},
  year     = {2024}
}

@inproceedings{gabriel2004openmp,
  address   = {Berlin, Heidelberg},
  author    = {
               Gabriel, Edgar and Fagg, Graham E. and Bosilca, George and others},
  booktitle = {
               Recent Advances in Parallel Virtual Machine and Message Passing
               Interface},
  pages     = {97--104},
  publisher = {Springer Berlin Heidelberg},
  title     = {
               {Open MPI: Goals, Concept, and Design of a Next Generation MPI
               Implementation}},
  year      = {2004}
}

@article{kennedy2001bayesian,
  author    = {Kennedy, Marc C and O'Hagan, Anthony},
  journal   = {Journal of the Royal Statistical Society: Series B (Statistical Methodology)},
  number    = {3},
  pages     = {425--464},
  publisher = {Wiley Online Library},
  title     = {Bayesian calibration of computer models},
  volume    = {63},
  year      = {2001}
}

@inproceedings{helmann2025gprat,
  address   = {Cham},
  author    = {Helmann, Maksim
               and Strack, Alexander
               and Pfl{\"u}ger, Dirk},
  booktitle = {Asynchronous Many-Task Systems and Applications},
  pages     = {83--94},
  publisher = {Springer Nature Switzerland},
  title     = {GPRat: Gaussian Process Regression with Asynchronous Tasks},
  year      = {2025}
}

@manual{mpi40,
    author = "{Message Passing Interface Forum}",
    title  = "{MPI}: A Message-Passing Interface Standard Version 5.0",
    url    = "https://www.mpi-forum.org/docs/mpi-5.0/mpi50-report.pdf",
    year   = 2025,
    month  = JUN,
    note   = {Accessed: 2026-01-10},
}

@misc{openblas,
  author       = {Xianyi Zhang and Qian Wang and Yunquan Zhang},
  title        = {{OpenBLAS}: An optimized BLAS library},
  howpublished = {\url{https://www.openblas.net}},
  year         = {2026},
  note   = {Accessed: 2026-01-10},
}

@article{Schuchart2025,
title={A Survey of Distributed Asynchronous Many-Task Models and Their Applications},
publisher={Institute of Electrical and Electronics Engineers (IEEE)},
author={Schuchart, Joseph and Diehl, Patrick and Bauer, Michael and others},
year={2025},
month=dec,
note={Preprint}
}

@Article{Thoman2018,
  author    = {Thoman, Peter and Dichev, Kiril and Heller, Thomas and others},
  journal   = {J. Supercomput.},
  title     = {A Taxonomy of Task-Based Parallel Programming Technologies for High-Performance Computing},
  year      = {2018},
  issn      = {0920-8542},
  month     = {04},
  number    = {4},
  pages     = {1422–1434},
  volume    = {74},
  publisher = {Kluwer Academic Publishers},
}

@ARTICLE{Boscila2013_parsec,
  author={Bosilca, George and Bouteiller, Aurelien and Danalis, Anthony and others},
  journal={Computing in Science \& Engineering}, 
  title={PaRSEC: Exploiting Heterogeneity to Enhance Scalability}, 
  year={2013},
  volume={15},
  number={6},
  pages={36-45},
}

@INPROCEEDINGS{Cao2022_parsec_cholesky_gp_cpu,
  author={Cao, Qinglei and Abdulah, Sameh and Alomairy, Rabab and others},
  booktitle={SC22: International Conference for High Performance Computing, Networking, Storage and Analysis}, 
  title={Reshaping Geostatistical Modeling and Prediction for Extreme-Scale Environmental Applications}, 
  year={2022},
  volume={},
  number={},
  pages={1-12}}

@ARTICLE{Abdulah2018_starpu_cholesky,
  author={Abdulah, Sameh and Ltaief, Hatem and Sun, Ying and others},
  journal={IEEE Transactions on Parallel and Distributed Systems}, 
  title={ExaGeoStat: A High Performance Unified Software for Geostatistics on Manycore Systems}, 
  year={2018},
  volume={29},
  number={12},
  pages={2771-2784},
}

@article{Augonnet2011_starpu,
  TITLE = {{StarPU: a unified platform for task scheduling on heterogeneous multicore architectures}},
  AUTHOR = {Augonnet, C{\'e}dric and Thibault, Samuel and Namyst, Raymond and others},
  JOURNAL = {{Concurrency and Computation: Practice and Experience}},
  PUBLISHER = {{Wiley}},
  SERIES = {Euro-Par 2009 best papers},
  VOLUME = {23},
  NUMBER = {2},
  PAGES = {187-198},
  YEAR = {2011},
}

@Article{Marcello2021,
  author    = {Dominic C Marcello and Sagiv Shiber and Orsola De~Marco and others},
  journal   = {Monthly Notices of the Royal Astronomical Society},
  title     = {{Octo-Tiger: a new, 3D hydrodynamic code for stellar mergers that uses HPX parallelization}},
  year      = {2021},
  month     = {04},
  number    = {4},
  pages     = {5345-5382},
  volume    = {504},
  publisher = {Oxford University Press ({OUP})},
}

@InProceedings{Strack2024_hpxfft,
author="Strack, Alexander
and Taylor, Christopher
and Diehl, Patrick
and others",
title="Experiences Porting Shared and Distributed Applications to Asynchronous Tasks: A Multidimensional FFT Case-Study",
booktitle="Asynchronous Many-Task Systems and Applications",
year="2024",
publisher="Springer Nature Switzerland",
address="Cham",
pages="111--122",
}

@misc{chatgpt,
  author       = {OpenAI},
  title        = {ChatGPT 5},
  howpublished = {\url{https://openai.com/chatgpt}},
  note         = {Accessed: 2026-01-10},
  year         = {2026}
}

@misc{grammarly,
  author       = {{Grammarly, Inc.}},
  title        = {Grammarly (Version 2.0)},
  howpublished = {\url{https://www.grammarly.com/}},
  note         = {Accessed: 2026-01-10},
  year         = {2026}
}

@misc{deepl,
  author       = {{DeepL SE}},
  title        = {DeepL Translator},
  howpublished = {\url{https://www.deepl.com/translator}},
  note         = {Accessed: 2026-01-10},
  year         = {2026}
}

@misc{gemini,
  author       = {{Google LLC}},
  title        = {Google Gemini 3},
  howpublished = {\url{https://gemini.google.com/}},
  note         = {Accessed: 2026-01-10},
  year         = {2026}
}

\end{document}